\documentstyle[aps,prl,twocolumn,epsf]{revtex}

\newcommand{\hal}[1]{{#1 \over 2}}

\def\Journal#1#2#3#4{{#1} {\bf #2}, #3 (#4)}

\def\NPA{{\em Nucl. Phys.} A}
\def\NPB{{\em Nucl. Phys.} B}
\def\PLB{{\em Phys. Lett.}  B}
\def\PRL{{\em Phys. Rev. Lett.}}
\def\PRD{{\em Phys. Rev.} D}

\def\PRV{{\em Phys. Rev.}}

\def\EPC{{\em Eur. Phys. J.} C}

\begin{document}

\draft
\title{Chiral Symmetry Realization for
 Even- and Odd-parity Baryon Resonances}

\author{D.\ Jido$^{(1)}$, T.\ Hatsuda$^{(1)}$, T.\ Kunihiro$^{(2)}$}
\address{$^{(1)}$ Department of Physics, Kyoto University,
 Kyoto, 606-8502, Japan}
\address{$^{(2)}$ Faculty of Science and Technology, 
Ryukoku University, Seta, Otsu-city, 520-2194, Japan}

\date{\today}

\maketitle
\begin{abstract}
  Baryon resonances with even and odd parity are collectively
  investigated from the viewpoint of chiral symmetry(ChS).  We propose
  a quartet scheme where $\Delta$'s and $N^{*}$'s with even and odd
  parity form a chiral multiplet.  This scheme gives parameter-free
  constraints on the baryon masses in the quartet, which are
  consistent with observed masses with spin $\hal{1},\hal{3},\hal{5}$. 
  The scheme also gives selection rules in the one-pion decay: The
  absence of the parity non-changing decay $N(1720) \rightarrow \pi
  \Delta(1232)$ is a typical example which should be confirmed
  experimentally to unravel the role of ChS in baryon resonances.

\end{abstract}

\pacs{12.39.Fe, 14.20.Gk}

  Chiral symmetry (ChS) and its
  dynamical breaking in quantum chromodynamics (QCD)
  are the key ingredients in low energy hadron
  dynamics.    
For instance, all hadrons can be classified in principle into
  some representation  of the chiral group
  $SU(N_{f})_{L} \times SU(N_{f})_{R} $, and 
  the interactions among
  hadrons are  strongly constrained by this  symmetry.

  There are two ways to realize ChS in effective
  low-energy Lagrangians;  non-linear and linear representations.
  The former has been extensively studied in the pion
  and nucleon sector
  and is summarized as  the celebrated chiral perturbation
  theory \cite{cp}. 
 The non-linear chiral
  transformations of the pion  and the nucleon are uniquely determined
  once we fix the parameterization of the coordinates of 
 the coset space $SU(N_{f})_{L} \times
  SU(N_{f})_{R} / SU(N_{f})_{V} $\cite{ccwz} and  the
  transformation of the nucleon under $SU(N_{f})_{V}$\cite{wein}. 
   On the other hand, in the linear representation,
  scalar mesons as chiral partners of the Nambu-Goldstone (NG) 
  bosons are introduced.
  Although such heavy mesons  do not
  allow systematic low energy expansion at zero temperature, 
   this representation is essential
  for studying  critical phenomena near the
  chiral phase transition where both the scalars and NG-bosons
  act as soft modes \cite{pw}.

  Then what about the baryons in the linear representation?
  The Gell-Mann L\'{e}vy sigma model \cite{GeL} is
  a first example where the nucleon transforms linearly
  both under the vector and the axial-vector transformations.
   DeTar and Kunihiro \cite{dk} 
generalized the model so that 
  $N_{+}$ (the nucleon) and its odd-parity partner $N_{-}$
  form a multiplet of  the chiral group. 
 (Note that, in  the non-linear representation, 
 different baryons do not form a multiplet
 by construction \cite{wein}.)
   A unique aspect  of their model is that the
  finite mass of  the nucleon 
  can be introduced in a chiral  invariant way, which
 opens a possibility that 
 even and odd parity nucleons may be degenerate
  with a non-vanishing mass in the  
 chirally symmetric phase. 

 In DeTar-Kunihiro's construction,
  $N_{\pm}$ are represented as a superposition of 
  $N_{1}$ and $N_{2}$ which are assigned to have
   opposite axial charges with each other.
  Subsequently this was  called  the ``mirror assignment"
 and distinguished from the 
 ``naive assignment" where $N_{1}$ and $N_2$ have
  the same axial change \cite{jnoh}: The two assignments are
  showed to   have 
 phenomenologically  distinguishable 
 predictions \cite{joh}.

  The purpose of this Letter is to develop the idea of the mirror 
  assignment in  baryon resonances with different parity
 ($P = \pm)$ and different isospin ($I =\hal{1}, \hal{3}$),
  and to explore how  ChS is realized in 
  the excited baryons. Achieving this purpose is tantamount to 
   constructing a linear sigma model
  in which both $\Delta_{\pm}$'s and 
  $N^{*}_{\pm}$'s 
  are incorporated for a given spin sector.  	
  (Here we call $N^*$ 
  ($\Delta$) as a resonance with $I= \hal{1}$ $(\hal{3})$,
  and the subscripts $\pm$ 
 denote their parity.)
 Thus we shall arrive at proposing a 
  {\em quartet scheme} in which $N^*_{+} , N^*_{-}, \Delta_{+}$ and 
  $\Delta_{-}$ form a chiral multiplet.  
  It will be shown that this quartet scheme
  is consistent with the observed baryon spectra without 
  fine-tuning of the model parameters.  We will also show some 
  evidence of this scheme in the decay pattern of the resonances.  
  Throughout the present Letter, we focus on $N_{f} = 2$, and 
  neglect the explicit breaking of ChS due to quark masses.

 To make the argument explicit,
 let us start with  $\Delta(1232)$ ($J^P=\hal{3}^{+}$)
 and its chiral partners. First of all, we need to choose the
  representation of $\Delta$ under 
 $SU(2)_{L} \times SU(2)_{R}$.
  The quark fields $q= q_{l} + q_{r}$ belong to 
  $(\hal{1},0) \oplus (0,\hal{1})$, where
   the first and second numbers in the parentheses refer
   to $SU(2)_{L}$ and $SU(2)_{R}$ representations, respectively.
   Therefore,  $(\hal{3},0) \oplus (0,\hal{3})$ and
   $(1,\hal{1}) \oplus (\hal{1},1)$ are the two candidates for
 $\Delta$;
 both of them  contain isospin  $I=\hal{3}$ and are constructed from 
 three quarks $[(\hal{1},0) \oplus (0,\hal{1}) ]^3$ \cite{cj}.
  Here, we choose 
  $(1,\hal{1}) \oplus (\hal{1},1)$ for $\Delta$, because 
 $\Delta$ is known to be 
   a strong resonance in $N$-$\pi$ system, and
    $N \times \pi = [(\hal{1},0) \oplus
   (0,\hal{1})] \times [(\hal{1},\hal{1})]$ does not contain 
 $(\hal{3},0) \oplus (0,\hal{3})$.
   In the quark basis, this representation may be schematically written as
 $  (1,\hal{1}) \oplus(\hal{1},1) =(q_{_L}q_{_L})_{I=1} q_{_R}
 \oplus  q_{_L} (q_{_R}q_{_R})_{I=1}$ where Lorentz and color indices
 are suppressed \cite{QSR}. Note that $ (1,\hal{1}) \oplus(\hal{1},1)$ 
 contains both $I=\hal{3}$ and
   $I=\hal{1}$ baryons, thus we utilize the latter to incorporate $N^*$.
 From now on, we do not
 consider the quark structure of $\Delta$ and $N^*$,
 and simply introduce elementary 
 Rarita-Schwinger (RS) fields for constructing 
 an  effective Lagrangian.

 To accommodate the parity partners of the baryon resonances, 
 let us define $\psi_{1}$ and
 $\psi_{2}$ as two independent  $J=\hal{3}$  RS fields 
  with even and odd parity, respectively.   The 
  Lorentz index $\mu =0,\ldots, 3$ for the RS fields is
 suppressed for brevity. We then  define the chiral 
 decomposition;  $\psi_{i} = \psi_{il} + \psi_{ir} $
 with $\gamma_5 \psi_{il, ir}= \mp \psi_{il, ir} $ ($i$=1,2).
  In the $J=\hal{3}$ chiral-quartet,  
$\psi_{1}$ and $\psi_{2}$ are mixed to form 
 four  resonances; $\Delta_{+}(P_{33})$, 
$\Delta_{-}(D_{33})$, $N^{*}_{+}(P_{13})$ and 
$N^{*}_{-}(D_{13})$.

   In the mirror assignment, 
 $\psi_{1l}$ and $\psi_{2 r}$ belong to $ (1,\hal{1}) $, 
  while $\psi_{1r}$ and $\psi_{2l}$ belong to $ (\hal{1},1) $ , 
 so that
   $\psi_1$ and $\psi_2$ have opposite axial charge.
 Thus, these
 fields have three indices, 
  $(\psi_{1,2})_{\alpha \beta}^{\gamma}$, with 
$ \alpha, \beta$ and $ \gamma $ take 1 or 2.  
  Here $(\alpha \beta)$
 is the
index for $I=1$
 triplet and $\gamma$ for $I=\hal{1}$ doublet.  
 Since $\psi$ is traceless
 for the triplet index $(\alpha \beta)$, it is convenient to 
 introduce a component field $(\psi_{i})^{A,\gamma}$ ($A=1,2, 3$ for triplet
 and $\gamma=1,2$ for doublet) as
\begin{eqnarray}
\label{comp}
(\psi_{1,2})_{\alpha \beta}^{\gamma} = 
\sum_{A=1,2,3} (\tau^A)_{\alpha \beta} (\psi_{1,2})^{A,\gamma} \ \ ,
\end{eqnarray}
where $\tau^{A}$ $(A = 1,2,3)$ is the $2 \times 2$ Pauli matrix.
 
 The transformation rules of $\psi_i$ under 
  $SU(2)_{L}\times SU(2)_{R}$  
   are then represented by
\begin{eqnarray}
\label{trapsi1l}
 (\tau^A)_{\alpha \beta} (\psi_{1l,2r})^{A,\gamma}
 & \rightarrow & (L \tau^A L^{\dagger})_{\alpha \beta}
 (R\psi_{1l,2r})^{A,\gamma}\ \ ,\\
\label{trapsi1r}
 (\tau^A)_{\alpha \beta} (\psi_{2l,1r})^{A,\gamma}
 & \rightarrow & (R \tau^A R^{\dagger})_{\alpha \beta}
 (L\psi_{2l,1r})^{A,\gamma}\ \ ,
 \end{eqnarray}
 where $L$ ($R$) corresponds to the $SU(2)_{L}$ ($SU(2)_{R}$) rotation.
 The meson field $M \equiv \sigma + i \vec{\pi}\cdot \vec{\tau}$
belongs to $(\hal{1},\hal{1})$ multiplet, and  obeys the 
 standard transformation rule,
$ M \rightarrow L M R^{\dagger}$.

Now let us construct the mass term and the Yukawa coupling of
  $\psi_{i}$ with $M$. 
  Here we consider only the simplest interaction which 
 has only single  $M$ without derivatives as in the case of the
  Gell-Mann-L\'{e}vy and DeTar-Kunihiro models. 
 It can be shown that the chiral invariance 
 under eq.'s.\ (\ref{trapsi1l},\ref{trapsi1r})  together with 
 parity and time-reversal invariance allow only three terms:
\begin{eqnarray}
    {\cal L}_{int} &=&  m_{0}\ 
    (\bar{\psi}^{A}_{2} \gamma_{5} \psi^{A}_{1} - \bar{\psi}^{A}_{1}
    \gamma_{5} \psi^{A}_{2})  \nonumber \\
    && + a\ \bar{\psi}_{1}^{A} \tau^{B} (\sigma  -i \vec{\pi}\cdot
    \vec{\tau}\gamma_{5}) \tau^{A} \psi_{1}^{B} \label{lag}\\
    && + b\ \bar{\psi}_{2}^{A}
    \tau^{B} (\sigma + i \vec{\pi}\cdot \vec{\tau} \gamma_{5})\tau^{A}
    \psi_{2}^{B}  \ \ , \nonumber
\end{eqnarray}
where  $m_{0}$, $a$ and  $b$ are free parameters not constrained by
 ChS.   This interaction for
  the $(1,\hal{1}) \oplus (\hal{1},1)$
 chiral-quartet is  a natural generalization
 of that for the $(\hal{1},0) \oplus (0,\hal{1})$
 chiral-doublet in \cite{dk}. 
 
 A short cut to obtain eq.\ (\ref{lag}) is
   to use $L M R^{\dagger}$ together
   with the rotated fields in the r.h.s.\ of 
   eq.(\ref{trapsi1l},\ref{trapsi1r}) 
  and to look for combinations
 in which $L$ and $R$ do not appear in the final expression.  
  Since $L$ and $R$ are independent transformation,  
    the indices related to the left (right)
 rotation must be always contracted with the left (right)
 rotation.  One of the chiral invariant  mass terms, for
  example, comes from the
  combination, ${\rm Tr}[(R \tau^A R^{\dagger})(R\tau^B R^{\dagger})] 
 [(\bar{\psi}^A_{1r}L^{\dagger})(L \psi^B_{2l})]$.
 Also, one of the Yukawa terms is obtained from
 $[(\bar{\psi}^A_{1l}R^{\dagger})(R \tau^B R^{\dagger})(R M^{\dagger}
 L^{\dagger})(L \tau^A L^{\dagger} )(L \psi^B_{1r})]$.

 As already mentioned,
  $\psi^{A,\gamma}_{i}$ contains both  $I=\hal{3}$ field
$\Delta_{i,M}$
$(M=\hal{3},\hal{1},-\hal{1},-\hal{3})$ and $I=\hal{1}$
field $N^{*}_{i,m}$ $(m=\hal{1},-\hal{1})$ which 
  are obtained by the following
  isospin
  decomposition:
 $\psi^{A,\gamma}_{i} = \sum_{M}(T^{A}_{3/2})_{\gamma M}\Delta_{i,M}$
 $+ \sum_{m}(T^{A}_{1/2})_{\gamma m} N^{*}_{i,m}$,
where the isospin projection matrices $T^{A}_{3/2}$ 
and $T^{A}_{1/2}$ are
defined through the Clebsh-Gordan coefficients, 
$(T^{A}_{3/2})_{\gamma M} = \sum_{r,\gamma^{\prime}} 
 (1r\hal{1}\gamma^{\prime} |\hal{3}M) 
 \epsilon_{r}^{A}\chi^{\gamma}_{\gamma^{\prime}}$ and 
$(T^{A}_{1/2})_{\gamma m} = \sum_{r,\gamma^{\prime}}
 (1r\hal{1}\gamma^{\prime} |\hal{1}m) 
 \epsilon_{r}^{A}\chi^{\gamma}_{\gamma^{\prime}}$. 
 $\vec{\epsilon}_{r}$ are vectors relating the $A=(1,2,3)$ basis
to $r=(+1,0,-1)$ basis, and
$\vec{\chi}_{\gamma^{\prime}}$ relates the $\gamma=(1,2)$ basis to
$\gamma^{\prime}=(\hal{1},-\hal{1})$ basis \cite{bw}. Their explicit
forms
 are
$\epsilon_{1} = -1/\sqrt{2}(1,i,0)$, $\epsilon_{0}=(0,0,1)$,
$\epsilon_{-1} = 1/\sqrt{2}(1,-i,0)$, $\chi_{1/2}=(1,0)$,
$\chi_{-1/2}=(0,1)$.

% \section{Masses of $\Delta$ and $N^{*}$}
With the invariant Lagrangian (\ref{lag}), we shall next show its 
phenomenological consequences on the 
masses of $\Delta$'s and $N^{*}$'s.  
 After the  spontaneously symmetry breaking 
$SU(2)_{L} \times SU(2)_{R} \rightarrow SU(2)_{V}$ due to the
 finite $\sigma$ condensate 
 $\langle \sigma \rangle \equiv \sigma_{0} >0 $, 
 the mass term in eq.(\ref{lag}) becomes
 \begin{eqnarray}
    {\cal L}_{m} &=& - (\bar{\Delta}_{1}, \bar{\Delta}_{2})
     \left(
     \begin{array}{cc}
         - 2 a \sigma_{0} &  \gamma_{5} m_{0}  \\
         -\gamma_{5} m_{0} & - 2 b \sigma_{0}
     \end{array} \right)
     \left(
     \begin{array}{c}
         \Delta_{1}  \\
         \Delta_{2}
     \end{array}\right) \nonumber \\ && -
      (\bar{N}^{*}_{1}, \bar{N}^{*}_{2})
     \left(
     \begin{array}{cc}
         a \sigma_{0} & \gamma_{5} m_{0}  \\
         -\gamma_{5} m_{0} &  b \sigma_{0}
     \end{array} \right)
     \left(
     \begin{array}{c}
         N^{*}_{1}  \\
         N^{*}_{2}
     \end{array}\right) \ .
 \end{eqnarray}

The physical bases $\Delta_{\pm}$ and $N_{\pm}^*$ diagonalizing
the mass matrices are given by 
 \begin{eqnarray*}
     \left(
     \begin{array}{c}
         \Delta_{+}  \\
         \Delta_{-}
     \end{array} \right) &=& {1 \over \sqrt{2\cosh \xi}} \left(
     \begin{array}{cc}
         e^{\xi / 2} & \gamma_{5} e^{-\xi / 2}  \\
         \gamma_{5} e^{-\xi / 2} &  -e^{\xi / 2}
     \end{array} \right) \left(
     \begin{array}{c}
         \Delta_{ 1}  \\
         \Delta_{2 }
     \end{array} \right)  \ ,
 \end{eqnarray*}
together with a similar formula for $ N^{*}_{\pm}$ with the replacement
   $\xi \rightarrow \eta$.
   The mixing angles $\xi$, $\eta$ are given by $\sinh \xi =
 -(a+b) \sigma_{0} /m_{0}$ and
  $\sinh \eta = (a+b) \sigma_{0}/ (2 m_{0})$. 
 These bases are chosen so that the masses of 
$\Delta$'s and $N^{*}$'s are all reduced to the chiral-invariant mass
 $m_0 >0$ when ChS is unbroken ($\sigma_0 =0$).
 
Thus we finally reach the mass formula,
\begin{eqnarray}
 \label{mass-f1}
    m_{\Delta_\pm} & = & \sqrt{(a+b)^{2} \sigma_{0}^{\, 2} 
    + m_{0}^{\,
    2}} \mp \sigma_{0}(a-b) ,\\
\label{mass-f2}
     m_{N^{*}_\pm} & = & \sqrt{(\hal{a+b})^{2} \sigma_{0}^{\, 2} +
    m_{0}^{\, 2}} \pm \hal{\sigma_{0}}(a-b) .
\end{eqnarray}
 Eq.'s\ (\ref{mass-f1},\ref{mass-f2}) shows 
that the spontaneous breaking of ChS
 lifts the degeneracy between 
 parity partners ($\Delta_{+}$ vs $\Delta_{-}$, and $N_{+}^*$ vs
$N_{-}^*$) 
 and the degeneracy  between  isospin states
  ($\Delta$ vs $N^{*}$)   simultaneously \cite{note2}. 

A remarkable consequence of our quartet scheme
 is  the following mass relations which hold irrespectively  of the
choice of the parameters ($m_0, a, b$):\\ 
%\begin{enumerate}    \item
1.\,  The ordering in parity-doublet of $N^*$ is
    always opposite to that of $\Delta$;
    \begin{eqnarray}
    \label{con1}
    {\rm sgn} \left[ m_{\Delta_{+}} - m_{\Delta_{-}} \right]
    = - \ {\rm sgn} \left[ m_{N_{+}^{*}} - m_{N_{-}^{*}} \right] \  .
    \end{eqnarray}
%\item 
2.\,     The mass difference between the two parity-doublets
  is fixed;
    \begin{equation}
    \label{con2}
  	\hal{1}(m_{\Delta_{-}}-m_{\Delta_{+}} )
      =  m_{N^{*}_{+}} - m_{N^{*}_{-}}\ .
    \end{equation}
% \item
3.\,   The averaged mass of the $\Delta$ parity-doublet is equal
    or heavier than that of $N^*$;
    \begin{equation}
 \label{con3}
	\hal{1}(m_{\Delta_{+}}+m_{\Delta_{-}}) \ge
	\hal{1}(m_{N^{*}_{+}} + m_{N^{*}_{-}}) \ .
    \end{equation}
%\end{enumerate}
 So far, we have considered only the case for $J=\hal{3}$. However,
 all the arguments and the mass relations above 
 hold for  the resonances with arbitrary spin as long as 
  $(1,\hal{1}) \oplus (\hal{1},1)$ chiral
 multiplets are concerned.

\begin{figure}[tbp]
   \centering
   \epsfxsize=8.0cm
   \epsfbox{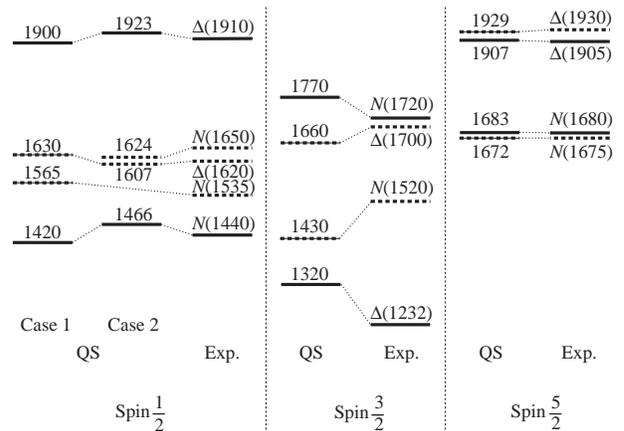}
    \caption{The quartet members with $J=\hal{1},\hal{3},\hal{5}$. The
    right (left)  hand side for each spin
    is the observed (quartet scheme) masses. The solid (dashed) lines denote 
    the even (odd) parity baryons. The reproduced masses in our scheme 
    agree with the experimental values within 10 \%.}
    \label{spec}
\end{figure}

 For the candidate of the quartets in the real world,
 we adopt
 the lightest baryons in each spin-parity among the established
 resonances with three or four stars in \cite{PDG}.
  $I=J=\hal{1}$ channel is, however, an exception since
  $N(940)$ is supposed to form  a
     $(\hal{1},0) \oplus (0,\hal{1})$ chiral doublet 
  with its parity partner which is either $N(1535)$ or
 $N(1650)$, or possibly their linear combination, in the mirror 
 assignment\cite{jnoh}.
 Therefore, we study two cases in $J = \hal{1}$ depending on
 whether we take $N(1535)$ (case 1) 
  or  $N(1650)$ (case 2) 
 as a $(1,\hal{1})\oplus (\hal{1},1) $ quartet member.
  In Fig.\ref{spec}, the observed resonances taken 
 from \cite{PDG} in the above criterion  are shown
   under the label ``exp" for each spin sector.

 The comparison between the mass relations in  the quartet
 scheme and  the experimental data are shown in the first
 three rows in Table
 \ref{tab.comp}.
 Parameter free constraints (\ref{con1}) and (\ref{con2})
  are well satisfied by the observed masses.    
 The constraint (\ref{con3}) 
 is well satisfied in $J=\hal{1}$ and $J=\hal{5}$ sectors, and is
 marginally satisfied in $J=\hal{3}$.

 If we have taken so called the ``naive assignment" where 
  $\psi_{1l,2l}$  belongs to $ (1,\hal{1}) $, 
 and $\psi_{1r,2r}$ belong to $ (\hal{1},1) $,
  the mass formula turns out to be the same with 
 eq.'s(\ref{mass-f1},\ref{mass-f2}) with $m_0=0$.
  This leads to a relation, 
 $m_{\Delta_{\pm}} = 2 m_{N^{*}_{\mp}}$, which is 
 in contradiction to the observed spectra in our
 criterion.
 This is  why 
 we have  not adopted the naive assignment in this Letter.

 Encouraged by the phenomenological success of the
 parameter free predictions of the
  mirror assignment,  we go one  step further
and determine  the three parameters  
 $m_{0}$, $a$ and  $b$  in each spin-sector. For this purpose,  
 we take the four observed masses  and $\sigma_{0}=f_{\pi}=93$ MeV
  and use the least square fit. 
 (For $J=\hal{3}$, we adopt $a=-b$ to satisfy the
 equality in eq.(\ref{con3}).) 
 Resultant parameters  are
 summarized in the last two rows  of Table \ref{tab.comp}.
  The  baryon masses in these parameters are also  shown under
 the label ``QS" in Fig.\ref{spec}.  They agree with the experimental
 data within $10$ percents.  
 
 $m_{0} \sim 1500$ MeV for $(1,\hal{1}) \oplus (\hal{1},1)$
 in Table \ref{tab.comp},
  which we obtained  irrespective of the spin,
   is considerably larger than $m_{0} = 270$ MeV
  for $(\hal{1},0) \oplus (0,\hal{1})$ \cite{dk}.
 Further investigation on the origin of  $m_0$ in QCD is necessary to
 understand  if these values as well as their difference  have
  physical implications. Also,  it is to be studied whether the baryonic 
 excitations with finite mass $m_0$ exist
  in the chiral restored phase using, e.g.,
 the lattice simulations.

% \section{$\pi$ couplings}
Let us return to the discussion of the $J=\hal{3}$ quartet
 and investigate the decay patterns 
 by the single pion emission obtained from eq.(\ref{lag}).  
 The interaction Lagrangian of $\pi$ and $\psi_{\pm}$ with 
 $a=-b=1.2$ is
 \begin{equation}
 \label{1pi}
     {\cal L}_{1 \pi} = 
     ( \bar{\psi}_{+}^{A}\, , \bar{\psi}_{-}^{A})
     \left(
     \begin{array}{cc}
 	0 & -a \\
 	a & 0
     \end{array}
     \right)
     \tau^{B}(i\vec{\pi}\cdot\vec{\tau})\tau^{A}
     \left(
     \begin{array}{c}
 	\psi_{+}^{B} \\
 	\psi_{-}^{B}
     \end{array}
     \right) \, ,
 \end{equation}
 where $\psi_{+}={1 \over \sqrt{2}}(\psi_{1}+\gamma_{5}\psi_{2})$
 and $\psi_{-}={1 \over \sqrt{2}}(\gamma_{5}\psi_{1}-\psi_{2})$.
 The mixing angles read
  $\xi$=$\eta$=0  due to $a+b=0$ (see Table 1).
 ${\cal L}_{1 \pi}$ has only the
 off-diagonal components in parity space: Therefore
 the parity
 non-changing couplings such as 
 $\pi \Delta_{\pm} N^{*}_{\pm}$,  $\pi \Delta_{\pm} \Delta_{\pm}$, 
 and  $\pi N^{*}_{\pm} N^{*}_{\pm}$
 are forbidden in the tree level of eq.(\ref{1pi}).

 Observed one-pion decay patterns are qualitatively consistent with the
 suppression of the $\pi\Delta_{+}N^{*}_{+}$ coupling.
 In fact, $N_{+}(1720)
 \rightarrow \pi \Delta_{+}(1232)$, although its phase space is
 large enough,  is insignificant or
 has not been shown to exist in the recent
 analysis of $\pi N$ scattering amplitudes \cite{man}.
  (The existence has been suggested in an old analysis of 
 $\pi N \rightarrow \pi\pi N$ though \cite{lon}.)
 On the other hand, $N_{-}(1520) \rightarrow \pi \Delta_{+}(1232)$ and
 $\Delta_{-}(1700) \rightarrow \pi \Delta_{+}(1232)$ in the $S$-wave
 channel, which are not suppressed in eq.(\ref{1pi}),
  have been seen with the partial decay rates $5\sim 12\%$ and
 $ 25\sim 50\%$, respectively\cite{PDG}.  
 The suppression of   $\pi \Delta_{\pm} \Delta_{\pm}$, 
 and  $\pi N^{*}_{\pm} N^{*}_{\pm}$ cannot be checked
 in the decays, but empirical studies of the $\pi N \rightarrow \pi 
 \pi N$ process\cite{arndt} seem to suggest that the $\pi 
 \Delta_{+}(1232)\Delta_{+}(1232)$ 
 coupling is less than half of the quark model 
 prediction given by $g_{\pi\Delta\Delta}=(4/5) g_{\pi NN}$ \cite{bw}.

 For $J=\hal{1}, \hal{5}$ sectors, similar analysis
 is not possible at present,  because of large
 uncertainties and/or the absence of experimental data
 for relevant decays.  
 More experimental data on the
 decays among quartet shown in Fig.1
 would be quite helpful for future theoretical studies.

 We note here that the
  selection rule discussed above may in principle
  be modified by chiral invariant terms not considered
 here, such as the terms containing
 derivatives as well as  multi $M$ fields.
  This is the situation similar to that for $g_{A}$ of the
 nucleon in  the linear sigma model,
  where the simplest Yukawa coupling in the
 tree level gives
 $g_A =1$ while the higher dimensional derivative coupling
 as well as quantum corrections could shift it to 1.25 \cite{lee}.
 Therefore, detailed studies with those terms should be also done
 in the future.

% \section{Conclusion}
In summary, we have investigated 
 baryon resonances with both parities from the viewpoint of chiral 
symmetry.
 We have constructed a linear sigma model in which $\Delta_{\pm}$'s and 
  $N^{*}_{\pm}$'s with a given spin are 
 assigned to be a representation $(1,\hal{1}) \oplus
(\hal{1},1)$ of the chiral $SU(2)_{L} \times SU(2)_{R}$ group.
Adopting the ``mirror assignment'' for the axial charge of baryons,
 we have arrived at
  {\em quartet scheme} where  $N^*_{+} , N^*_{-}, \Delta_{+}$ and 
  $\Delta_{-}$ form a chiral multiplet.
  We have  shown that the quartet scheme
  gives constraints not only on the baryon masses but also their 
 couplings;  it turns out that the constraints are
  consistent  with the observed baryon spectra.
 We have shown that experimental 
 confirmation of the absence of parity non-changing decay in
 $J=\hal{3}$ sector 
 such as  $N_{+}(1720) \rightarrow \pi \Delta_{+}(1232)$ together with the
 measurement of the decay patterns in 
 $J=\hal{1}, \hal{5}$ sectors is important to
test the quartet scheme and to explore the role of ChS in excited 
 baryons.

 D.\ J.\ is supported by Research Fellowships of the
Japan Society for the Promotion of Science for Young Scientists.
T. H. was partly supported by Grant-in-Aid for Scientific Research
 No. 10874042 of the Japanese Ministry of Education,
  Science and Culture.

%\end{document}

\begin{table}[h]
 \caption{Comparison between parameter free
 predictions of the  quartet scheme (QS)  and the observed data.
 Case 1 and case 2 in the $J=\hal{1}$ 
 sector stand for  the cases $N^{*}_{-}$=$N(1535)$ 
$N^{*}_{-}$=$N(1650)$, respectively. The last two
  rows are the parameters $m_0,a,b$
  determined from the experimental inputs.}
 \begin{center}
  \begin{tabular}{|c|c|c|c|c|c|}
       &  QS & \multicolumn{2}{c|}{$J=\hal{1}$} 
       &  $J=\hal{3}$  & $J=\hal{5}$  \\
       & & case 1 & case 2 & & \\
    \hline
    ${\rm sgn}\left( { m_{N_{+}^{*}} - m_{N_{-}^{*}} \over
                m_{\Delta_{+}}- m_{\Delta_{-}} } \right) $
       &  $-$ & $-$  & $-$   & $-$   &  $-$  \\
    \hline
    ${ m_{N_{+}^{*}} - m_{N_{-}^{*}} \over
                m_{\Delta_{+}}- m_{\Delta_{-}} } $
       & $-\hal{1}$   &  $-0.33$ &  $-0.72$  &  $-0.43$  &  $-0.2$  \\
    \hline
   ${ m_{N_{+}^{*}} + m_{N_{-}^{*}} \over
                m_{\Delta_{+}}+ m_{\Delta_{-}} } $
       & $ \le 1 $  &  0.84 & 0.88 & 1.1   &  0.87  \\
    \hline \hline
   \multicolumn{2}{|c|} {$m_0$ (MeV)} & 1380 & 1460 & 1540 & 1590 \\
   \hline
    \multicolumn{2}{|c|} {$(a, \ b)$}  & $(5.2, 6.6)$ & $(4.4, 6.1)$
   & $(1.2,-1.2)$ & $(5.8, 5.7)$ \\
  \end{tabular}
 \end{center}
 \label{tab.comp}
\end{table}

\end{document}